\documentclass[useAMS,usenatbib]{mn2e}

\usepackage{graphicx}
\usepackage{natbib}

\title[Mass profile slopes for dSphs]{Measuring the slopes of mass
profiles for dwarf spheroidals in triaxial CDM potentials}
\author[Laporte et al.]{
\parbox[t]{\textwidth}{
Chervin F. P. Laporte$^{1}$, Matthew G. Walker$^{2,3}$, Jorge
Pe\~narrubia$^{4,5}$}
\\
$^{1}$ Max Planck Institute for Astrophysics, Karl-Schwarzschild-Strasse
1, 85740 Garching, Germany\\
$^{2}$ Harvard-Smithsonian Center for Astrophysics, 60
Garden St., Cambridge, MA 02138, USA \\
$^{3}$ Hubble Fellow\\
$^{4}$ Institute for Astronomy, The University of Edinburgh, Royal
Observatory, Blackford Hill, Edinburgh EH9 3HJ, UK\\
$^{5}$ Ram\'on y Cajal Fellow, Instituto de Astrof\'isica de Andalucia
CSIC, Glorieta de la Astronom\'ia, 18008, Granada, Spain \\
}

\begin{document}
\date{}
\pagerange{\pageref{firstpage}--\pageref{lastpage}} \pubyear{2011}
\maketitle
\label{firstpage}
\begin{abstract}
We generate stellar distribution functions (DFs) in triaxial haloes
in order to examine the reliability of slopes $\Gamma\equiv \Delta {\rm log}
M / \Delta {\rm log} r$ inferred by applying mass estimators of the form 
$M\propto R_e\sigma^2$ (i.e. assuming spherical symmetry, where 
$R_e$ and $\sigma$ are luminous effective radius and global velocity 
dispersion, respectively) to two stellar sub-populations independently
tracing the same gravitational potential. The DFs take the form
$f(E)$, are dynamically stable, and are generated within triaxial potentials 
corresponding directly to subhaloes formed in cosmological 
dark-matter-only simulations of Milky Way and galaxy cluster haloes.  
Additionally, we consider the effect of different tracer number density
profiles (cuspy and cored) on the inferred slopes of mass profiles.  
For the isotropic DFs considered here, we find that halo triaxiality
tends to introduce an anti-correlation between $R_e$ and $\sigma$ when
estimated for a variety of viewing angles.  The net effect is a
negligible contribution to the systematic error associated with the
slope of the mass profile, which continues to 
be dominated by a bias toward greater overestimation of masses for
more-concentrated tracer populations.  We demonstrate that simple mass 
estimates for two distinct tracer populations can give reliable (and
cosmologically meaningful) lower limits for $\Gamma$, irrespective of the degree of triaxiality or 
shape of the tracer number density profile.

\end{abstract}

\section{Introduction}
The Milky Way's dwarf spheroidal (dSph) satellites include 
the most dark-matter-dominated galaxies known, with dynamical mass-to-light
ratios ranging from order $\sim 10$ to several hundreds in solar units \citep{Mateo1998}.
This makes dSphs objects of prime interest for studying the distribution of
dark matter in galaxies. dSphs lack atomic hydrogen; therefore methods for measuring dSph masses must rely on the kinematics of their pressure-supported stellar populations.  In the past decade, many techniques have been
developed with the goal of determining the internal mass distributions of dSphs: spherical Jeans
modelling \citep{Lokas2001,Strigari2006,Strigari2008, Koch2007, Battaglia2008, Walker2009, Wolf2010},
phase-space modelling \citep{Wilkinson2002, Amorisco2011},
the multiple stellar populations method \citep{Walker2011}, the use of the
virial theorem for spherical and constant flattened systems
\citep{Agnello2012} as well as axisymmetric
Jeans modelling \citep{Hayashi2012} and Schwarzschild modelling
\citep{Jardel2012, Breddels2012}.

Complicating most analyses is the fact that the inferred
dynamical mass is degenerate with the anisotropy of the velocity
dispersion tensor and the latter is poorly constrained by available 
line-of-sight velocity data.  While this degeneracy leaves the full mass profile 
underconstrained in a standard Jeans analysis
\citep{Strigari2006,Walker2009}, its relative weakness near
the halflight radius of the stellar tracer makes estimates
$M(R_e)\propto \kappa R_e\sigma^2$ (where $R_e$ and $\sigma$ are
luminous effective radius and global velocity dispersion,
respectively, and $\kappa$ is a constant) robust to various forms of
anisotropy and/or even to the shape of the mass profile \citep{Walker2009,Wolf2010}.

The presence of at least two chemo-dynamically distinct stellar
subpopulations in several dSphs \citep{Tolstoy2004, Battaglia2011} then provides a unique opportunity to measure the slopes of dSph mass
profiles, $\Gamma\equiv \Delta {\rm log} M / \Delta {\rm log} r$, directly by 
estimating $M(R_{e})$ at two different effective radii.  \citet[`WP11' hereafter]{Walker2011}
introduce a statistical method that uses estimates of stellar
positions, velocities and metallicities to estimate $R_e$ and $\sigma$
for each of two stellar subpopulations within the Fornax and
Sculptor dSphs, obtaining $\Gamma=2.61^{+0.43}_{-0.37}$ 
and $\Gamma=2.95^{+0.51}_{-0.39}$, respectively.  Taken at face value,
these measurements exclude, with significance $\sim 96\%$ and $\sim
99\%$, respectively, the \citet*['NFW' hereafter; $\Gamma \leq 2$ at all radii]{Navarro1997} 
profile 
that is often invoked to characterise
density profiles of cold dark matter (CDM) halos formed in
dissipationless cosmological simulations. WP11 tested their method against spherical dynamical models with various degrees of
anisotropy and found that mass estimators of the form $M(R_{e})\propto R_{e}\sigma^2$ systematically overestimate the enclosed
mass more strongly for tracers that are more deeply embedded (i.e.,
more concentrated) in their host haloes. This bias implies that slopes $\Gamma \equiv {\rm log}M / \Delta
{\rm log} r$ tend to be systematically {\it underestimated}, such that WP11's claim of
their quoted levels of NFW exlcusion were conservative.

However, despite the assumption of spherical symmetry that is common
to most dSph studies (exceptions include the axisymmetric
Schwarzschild analyses of \citealt{Jardel2012}
and the flattened models considered by \citealt{Agnello2012}), 
the composite stellar populations of real dSphs are clearly not
spherical.  The Milky Way's `classical' dSph satellites have
ellipticities ranging from $0.1 \la \epsilon\la 0.6$ \citep{Irwin1995}.
Furthermore, haloes formed in CDM cosmological simulations tend to be
triaxial \citep{Allgood2006,Vera-Ciro2011}.  Therefore,
insofar as CDM represents the null hypothesis regarding cosmological
structure formation, the relevance of inferences drawn from
spherically-symmetric analyses depends critically on their robustness
to axisymmetric and triaxial cases.

Here we test the slope measurements of WP11 for robustness against
non-spherical symmetry.  We exploit the fact that in a triaxial
potential, the energy is an integral of the motion and thus we can 
construct isotropic stellar distribution functions (DF) of the form
$f(E)$ even within triaxial N-body dark matter haloes. We use the
prescription presented by \cite{Laporte2013} to build stellar DFs with
various degrees of concentration within cosmological CDM haloes produced in
the Aquarius \citep{Springel2008} and Phoenix \citep{Gao2012} runs to cover a wide range of triaxiality parameters from Milky Way to cluster type environments.  
Section 2 discusses the numerical simulations and method used to
generate DFs.  Section 3 describes our use of samples from these DFs (projected along
various lines of sight) to examine systematic errors inherent to the
WP11 method for various forms of the tracer number density profiles.  
We discuss results and conclude in section~4.

\section{Numerical Methods}
\subsection{Dark matter haloes}

For the modelling of dSph dark matter haloes, we use the Aquarius simulations (see \cite{Springel2008} for details). This is a set of six high-resolution dark matter only simulations of the formation of Milky Way mass dark matter haloes in $\Lambda$CDM. In the level-2 resolution the particle mass is $\sim10^{4} \rm{M_{\odot}}$ and the softening length is $\epsilon= 65$ pc comoving. We extract a number of dark matter haloes in the mass range $10^{9}-10^{10} h^{-1} \rm{M_{\odot}}$, where $h=0.73$, using the subhalo finder {\sc subfind} \citep{Springel2001}.  The shape of the Aquarius subhaloes have axis ratios which increase with radius and which are mildly triaxial with axis ratios $<b/a> \sim 0.75$ and $<c/a> \sim 0.6$ at 1 kpc (Vera-Ciro private communication). We also complement our sample with subhaloes drawn from cluster simulations \citep{Gao2012} to bracket the range of possible triaxiality parameters for subhaloes in CDM and rescaled masses by a factor of 1000.

\subsection{Generating Tracers}
The weighting scheme used here was developed by \cite{Laporte2013} and is
a generalisation of that of \cite{Bullock2005} to triaxial systems. For details see \cite{Laporte2013} 
In short, in order to generate a luminous stellar
profile, we take each simulation particles of energy
$E=\frac{1}{2}v^{2}+\Phi$ to simultaneously represent dark matter and
stars in diferent amounts through the weight function
$\omega(E)=\frac{N_{*}(E)}{N(E)}=\frac{f_{*}(E)g(E)}{f(E)g(E)}$, where $N$
is the differential energy distribution, $g$ is the density of states and
asterisks denote stellar quantities. One generates $f_{*}(E)$ through
specifying the target number density profile $\nu=\nu(r)$ and using the
Eddington formula with an additional approximation to deal with the
multivalued behaviour of $\Phi=\Phi(r)$ with spherical radius. In this
way, the method creates a stellar profile which retains
contours of the flattening of the total potential. 

Figure 1 displays projected number densities and line of sight velocity dispersion
profiles obtained by sampling random projections of DFs (for two
different stellar number density profiles and different concentrations) calculated
using the machinery described above.  We check for equilibrium by tracking the 
stellar DFs forward in time for a period of 150 Myr, during which the stellar 
density profiles remain stable.  

\begin{figure}
\includegraphics[width=0.5\textwidth,trim=0mm 0mm 0mm
0mm,clip]{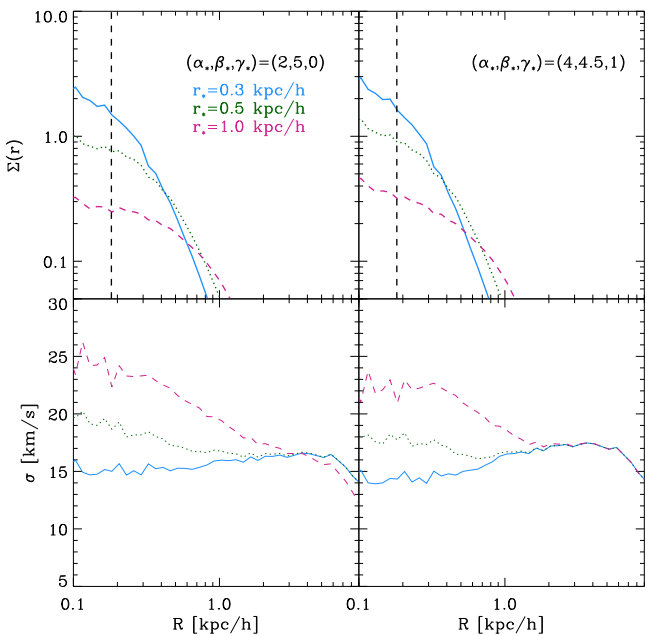}
\caption{Projected number density and line of sigh velocity dispersion
profiles for one Aquarius subhalo using the present weighting scheme. {\it
Top left:} Plummer profiles with $r*=0.3,0.5,1.0 h^{-1} \rm{kpc}$, {Top
right}: $(\alpha_{*},\beta_{*},\gamma_{*})=(4,4.5,1)$ profiles with
$r_{*}=0.3,0.5,1.0 h^{-1} \rm{kpc}$. The vertical dotted line marks the point where $r=2.8\epsilon$.}
\end{figure}

WP11 originally tested their method using models in which stellar
populations trace dark matter potentials characterised by generalised
 \citet[see also \citealt{Zhao1996}]{Hernquist1990} profiles:

\begin{equation}
\nu_{*}(r)=\nu_{0}\left(\frac{r}{R_{e}}\right)^{-\gamma_{*}} \left[1+
\left(\frac{r}{R_{e}}\right)^{\alpha_{*}}
\right]^{(\gamma_{*}-\beta_{*})/\alpha_{*}}
\end{equation}
and
\begin{equation}
\rho_{dm}(r)=\rho_{0}\left(\frac{r}{r_{dm}}\right)^{-\gamma_{dm}} \left[1+
\left(\frac{r}{r_{dm}}\right)^{\alpha_{dm}}
\right]^{(\gamma_{dm}-\beta_{dm})/\alpha_{dm}}.
\end{equation}
For the stellar number densities, WP11 considered Plummer profiles,
$(\alpha_{*},\beta_{*},\gamma_{*})=(2,5,0)$, which provide good fits to
dSph surface brightness profiles \citep{Irwin1995,McConnachie2006} and has the virtue of
depending on a single parameter, the projected halflight radius $R_e$.  They also
considered alternative profiles that retain a luminous core ($\gamma_{*}=0$) but
fall off more slowly/quickly at large radius than do Plummer profiles\footnote{They actually considered models with $\gamma_{*}=0.1$ because models $\gamma=0$ have $f(E)<0$ in some regions. In the simulations the resolution limit already prevents this from happening for our models.},
with $(\alpha_{*},\beta_{*},\gamma_{*})=(2,4,0)$ and $(2,6,0)$, respectively. \cite{Strigari2010} have shown that cuspy tracer number density profiles
provide a good match to the observed surface brightness and velocity
dispersion profiles of the composite stellar populations in dSphs.  
At a fixed half-light radius, a cuspy tracer
component would have a lower velocity dispersion than would its cored
counterpart.  In order to test for sensitivity to the inner profile of
the tracer components, here we consider models with stellar cusps
$(\alpha_{*},\beta_{*},\gamma_{*})=(4,4.5,1)$ as well as cored Plummer profiles
with $(2,5,0)$.  

\section{Mass modelling: multi-component method}
The presence of multiple stellar populations in some dSphs
enables the observer to estimate enclosed masses at two different half-light radii
in the same potential.  Testing their method on DFs drawn from
spherically-symmetric models with cored light profiles, WP11 find that masses tend to be
over-estimated more strongly for more-concentrated stellar
populations.  As a result, the slope $\Gamma$ tends to be
underestimated, providing conservative lower limits on the true
slope.  We now use our models $f_{*}(E)$ to test whether this behavior holds for
the case of triaxial haloes and/or when the tracer number density
profiles are cusped instead of cored.  

\subsection{The bias in the WP mass-estimator: systematics}

After calculating DFs as described above, we project each model along 100 random lines of
sight uniformly sampled on a sphere. For each projection angle, we then
calculate the half-light radius $R_{e}$ of each population.  In order to mimic the WP11 method, we estimate $R_{e}$ by $\chi^{2}$-fitting a Plummer profile to the
tracers. The mass enclosed within $R_{e}$ is then

\begin{equation}
M(R_{e})\propto R_{e}
\frac{\Sigma_{i=0}^{N}w_{i}(v_{i}-\bar{v})^2}{\Sigma_{i=0}^{N} w_{i} }
\propto R_{e} \sigma^2,
\end{equation}
where $w_{i}$ are the N-body particle weights, and $\sigma$ is the global
velocity dispersion of the tracers. The slope is then calculated as
$\Gamma=\frac{log(M_{1}/M_{2})}{log(r_{1}/r_{2})}$. 

In order to check whether the WP11 method continues to give
conservative limits, Figure 2 displays distributions of the bias
$E[\Gamma]=\Gamma_{est}-\Gamma_{true}$ over all randomly-chosen
viewing angles.  In nearly all cases the estimated slope is smaller
than the true slope, such that the estimated slopes continue to
represent conservative lower limits.  This behavior holds regardless
of the degree of triaxiality and/or whether the light profile is
cusped or cored.

We emphasise that the velocity dispersion that enters the WP11 mass
estimator, $ M \propto R_{e}\sigma^2$ refers to the global dispersion of
all stars in a given stellar sub-population.  Recently, \citet[`K12' hereafter]{Lucio2012}
have found that use of a different mass estimator - one that refers to the velocity
dispersion only of stars inside the half-light radius - would give less
reliable limits on $\Gamma$, particularly when triaxiality is
present.  We confirm this result using our
own DFs (Figure 3): indeed, when velocity dispersions are estimated using only
stars inside $R_{e}$ of their respective subpopulation, the estimated slopes have large scatter about the
true values and do not constitute reliable lower limits.  

In addition to stellar number densities used in studying dwarf spheroidals,
we also show results from additional tests for which we adopted the
Jaffe profile, $(\alpha_{*},\beta_{*},\gamma_{*}) =(2,0,2)$, which can be used to
model ellipticals and which has steep stellar cusp
($\gamma_{*}=2$). In this case, we determine the
half-light radius by fitting a de Vaucouleurs profile.  For these
cases with steep stellar cusps, we find that that WP11's method
becomes unreliable when the stellar populations are highly
concentrated (top panel in Fig. 2); however, for
sufficiently extended stellar populations the method still recovers a
conservative (i.e., biased towards low values) estimate of the slope of the underlying mass profile, albeit
with a more prominent tail towards positive values. 

\begin{figure}
\includegraphics[width=0.5\textwidth,trim=0mm 0mm 0mm
0mm,clip]{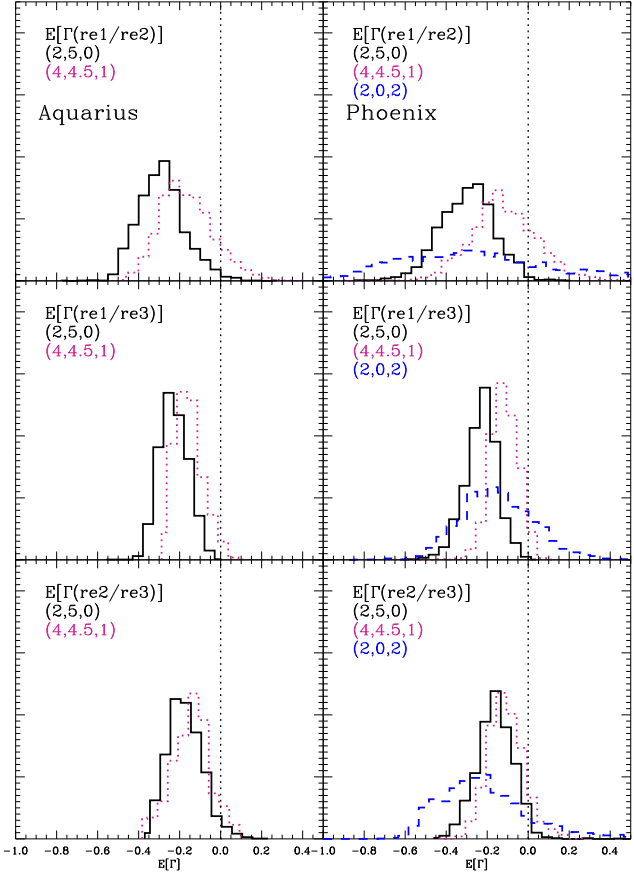}
\caption{The stacked bias distribution in the slope determination of live
N-body dark matter haloes (all observed through 100 different random lines
of sights). The vertical line marks the point where
$\Gamma_{est}-\Gamma_{true}$ is zero. 
{\it Left hand:} Results for Aquarius subhaloes for Plummer and
$(4,4.5,1)$ profiles (in black and dotted magenta respectively) {\it Right Hand:}: Phoenix
rescaled subhaloes for Plummer, $(4,4.5,1)$ and Jaffe profiles (in black, dotted magenta and dashed blue
respectively). The half-light radii of the
stellar populations are determined through fitting a Plummer profile to
the number density profile (as assumed in WP11). $(r_{e1},r_{e2},r_{e3})=(0.3,0.5,1.0) h^{-1} kpc$.}
\end{figure}
\begin{figure}
\includegraphics[width=0.5\textwidth,trim=0mm 0mm 0mm
0mm,clip]{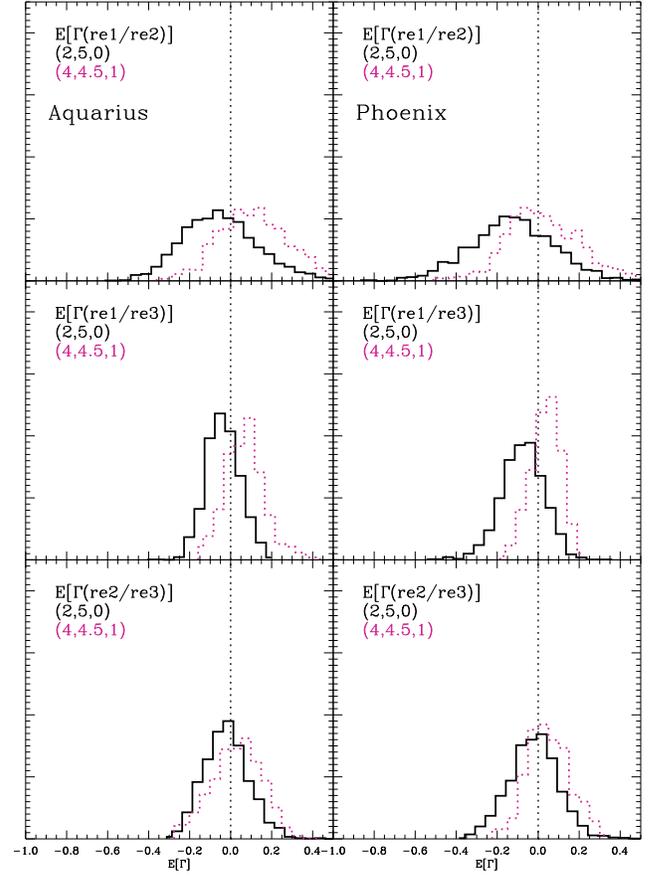}
\caption{Same as figure 2 but showing results derived using the method
used in Kowalczyk et al. 2012. Clearly losing the kinematic information
outside the half-light radius of the tracer makes the estimator highly
unreliable. This method is not that used in WP11.}
\end{figure}

\subsection{Why triaxality does not matter so much?}
We can understand the relative insensitivity of the WP11 method to
triaxiality by considering the coupling of estimated quantities $R_{e}$
and $\sigma$ with respect to projection angle.  Let us
 rotate an individual halo in the frame of its body axes (as evaluated
 at a radius of $1 h^{-1} \rm{kpc}$) such that the major axis lies on the
$x$-axis and the minor axis lies on the $y$-axis.  We then observe it along
different polar angles in the $x-y$ plane and estimate the half-light
radius and velocity dispersion via the same $\chi^{2}$ fitting
procedure used above. We notice that when the velocity dispersion is
large (along the major axis) the estimated value of $R_{e}$ is at its
minimum value and vice versa (Figure 4).  This anti-correlation of $R_{e}$ and
$\sigma$ tends to cancel the effects of triaxiality on the mass
estimator. Therefore the slope $\Gamma$ will be less sensitive because at a fixed angle $\theta$ any bias in $M(R_{e})$ will cancel out in the estimate of $\Gamma$. This is why the biases we recover in Figure 2 are similar to those found by WP11 for spherically symmetric models. The fluctuations in the
mass estimates due to triaxiality vary from 10 to 20 percent depending on the embededness of the tracer population.

\begin{figure}
\includegraphics[width=0.5\textwidth,trim=0mm 0mm 0mm
0mm,clip]{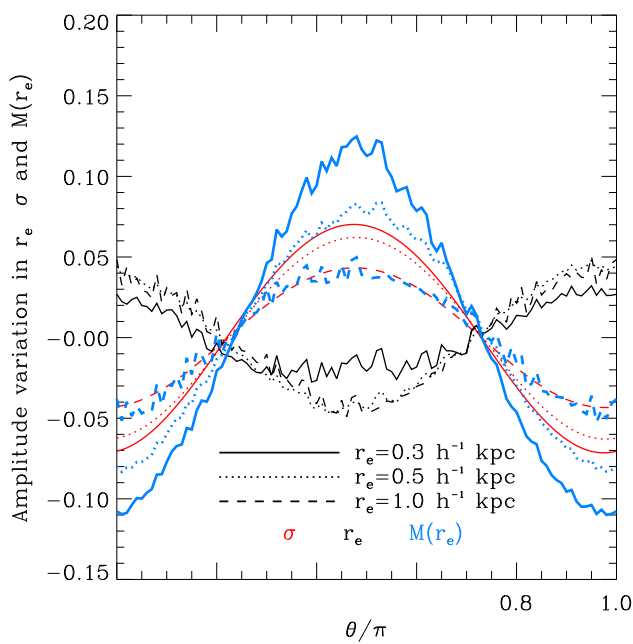}
\caption{Amplitude of the variations in $R_e$ (black), $\sigma$ (red) and
$M(r_{e})$ (blue) for different stellar populations (different line
styles) for an example halo from Aquarius as a function of angle $\theta$,
the polar angle in the plane of the major and minor axes of the halo
evaluated at $1 h^{-1} \rm{kpc}$.  The anti-correlation in the behaviour of $R_e$ and $\sigma$ creates almost a cancellation and weak variations in $M(R_e)$. We also see that variations are greater the more embedded the stellar system similarly to the mass bias observed in WP11. This in turn explains why the slopes estimates are still reliable under our $f(E)$ models in triaxial potentials.}
\end{figure}

\section{Discussion and conclusions}

We have presented families of isotropic distribution functions of the form $f(E)$ in
triaxial potentials extracted from dark-matter-only simulations. These
span a range of dark matter density profiles for which we have tested the method of
\cite{Walker2011}. Our tests show that the method is
generally able to place conservative limits on slopes of mass
profiles, even when the light profiles have NFW-like cusps as advocated
by \citet{Strigari2010}. Thus, we conclude that triaxiality has little impact on published
analyses of dSph stellar kinematics that assume spherical symmetry.
The reason is that $R_{e}$ and $\sigma$ are anti-correlated over the
range of projection angles, effectively cancelling the effects of triaxiality.
However, we have found that the WP11 method can break down if the
stellar tracers are highly concentrated and have steeply cusped number
density profiles, e.g., the Jaffe profiles examined in Section 3.1. Some of the haloes which were identified by {\sc subfind} are strongly
stripped and the tracer may not be entirely in equilibrium. However, using
those models, we were still able to recover successful limits on the slope
of the dark matter density profiles. This suggests that tidal
stripping does not unduly impact the results of WP11. 

Recently, a similar study on the same subject has been carried out
by K12. Our work differs in three aspects:
\begin{enumerate}
\item We consider haloes which form within a $\Lambda$CDM cosmological context. K12 have considered
spherical models which get tidally stirred under a static potential.
\item Our models do not have rotation. Many galaxies in K12 still retain rotation, which is not observed in dSphs.
\item K12 do not test the robustness of WP11 to triaxality, but show that a mass estimator based on the velocity dispersion within the half-light radius of a tracer can misinterpret the true value of the slope of the total mass profile. We confirm their result in Figure 3.
\end{enumerate}

Finally, we note that given a density profile, there exists many possible
velocity dispersion profiles which may be consistent with the observed
data (allowing for anisotropy). However, WP11 showed this is not an issue
for their method under anisotropic Ossipkov-Merritt models but also those
with constant anisotropy. Combined with the results of our current study, the WP11 method seems to 
be robust to both anisotropy and halo triaxiality.


\section*{Acknowledgments}
CFPL thanks Simon White for useful discussions, Mark Gieles and the IoA where early discussions began. MGW and JGP thank the MPA for its hospitality during their visit. The authors thank the Virgo Consortium for making their data available for this study. CFPL is supported by the Marie Curie Initial Training Network CosmoComp (PITN-GA-2009-238356). MGW is supported by NASA through 
Hubble Fellowship grant HST-HF-51283.01-A, awarded by the Space Telescope 
Science Institute, which is operated by the Association of Universities for 
Research in Astronomy, Inc., for NASA, under contract NAS5-26555. 

\bibliographystyle{mn2e}
\bibliography{master2.bib}{}
\label{lastpage}
\end{document}